\RequirePackage{fix-cm}
\documentclass[reqno,intlimits]{amsproc}
\usepackage{fixltx2e}

\usepackage{amssymb}

\usepackage[citation-order,nobysame]{amsrefs}
\usepackage{xyzbib}

\usepackage{pstricks,pst-node}

\usepackage{hyperref}

\usepackage[foot]{amsaddr}

\numberwithin{equation}{section}

\newcommand\lmove{\hspace{-.25in}}

\makeatletter
\def\@setemails{%
   \lmove\mbox{{\itshape E-mails}:\space}{\ttfamily\emails}.
 }
\makeatother
\DeclareSymbolFont{lsymb}{U}{euex}{m}{n}
\DeclareMathSymbol{\intop}{\mathop}{lsymb}{"52}
\DeclareMathSymbol{\ointop}{\mathop}{lsymb}{"48}
\DeclareMathSymbol{\smallint}{\mathop}{lsymb}{"52}

\renewcommand{\leq}{\leqslant}

\DeclareMathOperator*{\Asym}{\textrm{Asym}}

\newcommand{\rmi}{\mathrm{i}}
\newcommand{\rmd}{\mathrm{d}}
\newcommand{\eps}{\varepsilon}

\newcommand{\ot}{\mathop{\otimes}}
\newcommand{\bra}[1]{\left\langle #1 \right\rvert}
\newcommand{\ket}[1]{\left\lvert #1 \right\rangle}

\newcommand{\Ztop}[1]{Z^{\mathrm{top}}_{#1}}
\newcommand{\Zbot}[1]{Z^{\mathrm{bot}}_{#1}}
\begin{document}

\title{On the problem of calculation of correlation functions
in the six-vertex model with domain wall boundary conditions}

\author{F. Colomo$^{1)}$}
\address{\lmove$^{1)}$INFN, Sezione di Firenze,
Via G. Sansone 1, 50019 Sesto Fiorentino (FI), Italy}
\email{colomo@fi.infn.it}
\author{A. G. Pronko$^{2)}$}
\address{\lmove$^{2)}$Saint Petersburg Department of V.~A.~Steklov
Mathematical Institute,
Russian Academy of Sciences,
Fontanka 27, 191023 Saint Petersburg, Russia}
\email{agp@pdmi.ras.ru}

\begin{abstract}
The problem of calculation of correlation functions in the
six-vertex model with domain wall boundary conditions is addressed
by considering a particular nonlocal correlation function, called row configuration
probability. This correlation function can be used as building block for
computing various (both local and nonlocal) correlation functions in the
model. The row configuration probability is calculated using the quantum
inverse scattering method; the final result is given in terms of a multiple
integral. The connection with the
emptiness formation probability, another nonlocal correlation function
which was computed elsewhere using similar methods, is also discussed.
\end{abstract}

\maketitle

%%%%%%%%%%%%%%%%%%%%%%%%%%%%%%%%%%%%%%%%%%%%%%%%%%%%%%%%%%%%%%%%%%%%
\section{Introduction}

One of the most fundamental problems in the theory of integrable models
is the exact calculation of correlation functions \cite{KBI-93}. In
recent years, there has been an increasing interest (motivated by various
mathematical and physical applications) in obtaining exact results for
correlation functions of  statistical mechanics models defined on
finite lattices and with fixed boundary conditions. Because of the lack
of translational  invariance, the systematic computation of  correlation
functions for these models represents a difficult problem.

An important example of such a model is the six-vertex model with domain
wall boundary conditions \cite{K-82}. The partition function of the model
on the finite lattice is given exactly in terms of certain determinant
\cite{I-87,ICK-92}. This formula, known as Izergin-Korepin
determinant formula, turned out a powerful tool in proving important
combinatorial results. The current interest in the model is mostly
motivated by occurrence of the phase separation phenomena (see
\cite{CP-08,CP-09} and references therein).

Some progress in the calculation of the correlation functions
of the six-vertex model with domain wall boundary
conditions has been achieved when correlations are considered
near the boundaries \cites{BKZ-02,BPZ-02,FP-04,CP-05c}. An example of
correlation function which can be computed away from the boundary is the
so-called emptiness formation probability \cite{CP-07b}. Some
generalisations have been considered recently in \cite{M-11}. However the problem
of a systematic treatment of correlation functions, especially when
correlations are considered away from the boundaries, is still far from being
solved.

To address this problem, in the present paper we introduce a particular non-local
correlation function, called row configuration probability. This correlation
function describes the probability of observing a given configuration of
arrows on the vertical edges located between two consecutive horizontal lines
of the square lattice. The row configuration probability can be used as
building a block to compute other (both local and non-local) correlation
functions. In particular, it is closely related to the emptiness formation
probability.

The row configuration probability besides being interesting for evaluation of
other correlation functions, is also interesting on its own right: there are
analogue in the context of phase separation for dimers models \cite{DR-09},
and of enumerative combinatorics \cite{FR-10}.

To compute the row configuration probability we consider the inhomogeneous
version of the model and formulate it in the framework of the quantum inverse
scattering method (QISM) \cite{TF-79} (for a survey, see \cite{KBI-93}). We
use the fact that the row configuration probability can be represented as a
product of two factors. For computing the first factor we use a side result
of paper \cites{IKR-87}, while for the second one we use the technique
developed in papers \cites{BPZ-02,CP-05c,CP-07b}. For the homogeneous model
both factors are represented in terms of multiple integrals.

To demonstrate how these results can be used for computing other correlation
functions, we discuss here the connection with the emptiness formation
probability. The computation is based on performing certain sums
and integrals, and making use of identities involving antisymmetrisation of
multi-variable functions.

The paper is organized as follows. In the next section after
recalling basic facts about the model we
set up the considered problem in terms of QISM objects.
The derivation of the row configuration
probability is given in sections 3 and 4.
The relation with the emptiness formation probability is discussed
in section 5.

%%%%%%%%%%%%%%%%%%%%%%%%%%%%%%%%%%%%%%%%%%%%%%%%%%%%%%%%%%%%%%%%%%%%
\section{Six-vertex model, domain wall boundary conditions, and
row configuration probability}

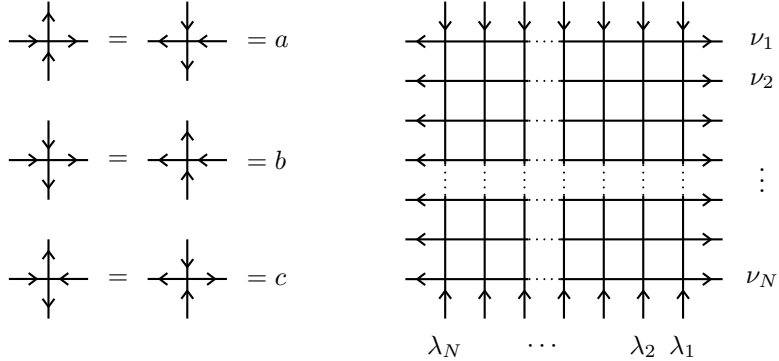
\begin{figure}
\centering
\psset{unit=15pt,dotsep=2pt}
\newcommand{\arr}{\lput{:U}{\begin{pspicture}(0,0)
\psline(0,0.15)(.2,0)(0,-.15) \end{pspicture}}}

%\begin{document}

\begin{pspicture}(0,-1)(19,8)
%\psframe(0,-1)(19,8)
\rput(0,6){
\pcline(0,1)(1,1)\arr \pcline(1,1)(2,1)\arr
\pcline(1,0)(1,1)\arr \pcline(1,1)(1,2)\arr}
\rput(3.5,6){
\pcline(1,1)(0,1)\arr \pcline(2,1)(1,1)\arr
\pcline(1,2)(1,1)\arr \pcline(1,1)(1,0)\arr}
\rput[l](2.5,7){$=$}
\rput[l](6,7){$=a$}
\rput(0,3){
\pcline(0,1)(1,1)\arr \pcline(1,1)(2,1)\arr
\pcline(1,2)(1,1)\arr \pcline(1,1)(1,0)\arr}
\rput(3.5,3){
\pcline(2,1)(1,1)\arr \pcline(1,1)(0,1)\arr
\pcline(1,0)(1,1)\arr \pcline(1,1)(1,2)\arr}
\rput[l](2.5,4){$=$}
\rput[l](6,4){$=b$}
\rput(0,0){
\pcline(0,1)(1,1)\arr \pcline(2,1)(1,1)\arr
\pcline(1,1)(1,0)\arr \pcline(1,1)(1,2)\arr}
\rput(3.5,0){
\pcline(1,1)(2,1)\arr \pcline(1,1)(0,1)\arr
\pcline(1,2)(1,1)\arr \pcline(1,0)(1,1)\arr}
\rput[l](2.5,1){$=$}
\rput[l](6,1){$=c$}
\rput(10,0){
\multirput(1,0)(1,0){7}{\pcline(0,0)(0,1)\arr \pcline(0,8)(0,7)\arr}
\multirput(0,1)(0,1){7}{\pcline(1,0)(0,0)\arr \pcline(7,0)(8,0)\arr}
\multirput(1,1)(0,1){7}{\psline(0,0)(2.1,0)\psline[linestyle=dotted](2.1,0)(2.9,0)\psline(2.9,0)(6,0)}
\multirput(1,1)(1,0){7}{\psline(0,0)(0,2.1)\psline[linestyle=dotted](0,2.1)(0,2.9)\psline(0,2.9)(0,6)}
\rput(9,1){$\nu_N$}
\rput(9,3.7){$\vdots$}
\rput(9,6){$\nu_2$}
\rput(9,7){$\nu_1$}
\rput(1,-.7){$\lambda_N$}
\rput(3.5,-.7){$\cdots$}
\rput(6,-.7){$\lambda_2$}
\rput(7,-.7){$\lambda_1$}
}
\end{pspicture}
%\end{document}
\caption{The six vertices and their weights (left), and
$N$-by-$N$ square lattice with domain wall boundary conditions (right).}
\label{fig.DWBC}
\end{figure}

We consider the six-vertex model on a square lattice formed by intersection
of $N$ horizontal and $N$ vertical lines (an $N\times N$ lattice), with
special fixed boundary conditions called domain wall boundary conditions.
Recall that the six-vertex model is a model in which local states are arrows
pointing along edges of the lattices; the allowed arrow configurations are
subject to the `ice-rule': each vertex should have the same number of
incoming and outgoing arrows. The Boltzmann weights are assigned to the six
possible vertex configurations of arrows allowed by the ice-rule, and in the
model invariant under reversal of all arrows there are three different
Boltzmann weights, usually denoted $a$, $b$, and $c$. The domain wall
boundary conditions mean that all arrows on the left and right boundaries are
outgoing while all arrows on the top and bottom boundaries are incoming, see
figure \ref{fig.DWBC}.

To use QISM in calculations we consider the inhomogeneous version
of the model, in which the weights of the vertex being at the intersection of
$\alpha$th vertical line (enumerated from the right) and $k$th horizontal line (enumerated from the top) are
$a_{\alpha, k}=a(\lambda_\alpha,\nu_k)$, $b_{\alpha, k}=b(\lambda_\alpha,\nu_k)$, and
$c_{\alpha, k}=c$, where
\begin{equation}\label{abc}
a(\lambda,\nu)=\sin(\lambda-\nu+\eta),\qquad
b(\lambda,\nu)=\sin(\lambda-\nu-\eta),\qquad
c=\sin2\eta.
\end{equation}
The parameters $\lambda_1,\dots,\lambda_N$ are assumed to be all different;
the same is assumed about $\nu_1,\dots,\nu_N$. The parameter
\begin{equation}\label{Delta}
\Delta=\frac{a_{\alpha, k}^2+b_{\alpha, k}^2-c_{\alpha, k}^2}{2 a_{\alpha, k}b_{\alpha, k}}=\cos2\eta
\end{equation}
takes the same value for all vertices, that ensures integrability \cite{B-82}.
The partition function is defined as follows
\begin{equation}%\label{}
Z_N= \sum_{\mathcal{C}}^{}
\prod_{\alpha,k=1}^{N}w_{\alpha, k}(C),
\end{equation}
where $w_{\alpha, k}(C)$ takes values $w_{\alpha, k}(C)= a_{\alpha
k},b_{\alpha k},c_{\alpha k}$, depending on the configuration $C$. Clearly,
$Z_N=Z_N(\lambda_1,\dots,\lambda_N;\nu_1,\dots,\nu_N)$ where
$\lambda_1,\dots,\lambda_N$ and $\nu_1,\dots,\nu_N$ can be regarded as
`variables'; parameter $\eta$ has the meaning of a `coupling constant' and it
is often omitted in the notations. After QISM calculations, the homogeneous
model quantities (e.g., partition function) can be obtained from the
inhomogeneous ones upon setting $\lambda_\alpha=\lambda$ ($\alpha=1,\dots,N$)
and $\nu_k=\nu$ ($k=1,\dots,N$), where, with no loss of generality, one can
further put $\nu=0$, see \eqref{abc}. We shall refer to this procedure as
homogeneous limit.

We now define the main objects of QISM in relation to the model.
First, let us consider vector space $\mathbb{C}^2$ and denote its basis
vectors as the spin-up and spin-down states
\begin{equation}%\label{}
\ket{\uparrow}=
\begin{pmatrix}
1 \\ 0
\end{pmatrix},\qquad
\ket{\downarrow}=
\begin{pmatrix}
0 \\ 1
\end{pmatrix}.
\end{equation}
To each horizontal and vertical line of the lattice
we associate vector space $\mathbb{C}^2$.
We also use the convention that upward and right arrows correspond to
the `spin up'  state while downward and left arrows correspond
to the  `spin down' state.

Next, to each vertex being intersection of the
$\alpha$th vertical line and the $k$th horizontal line
we associate the
operator $L_{\alpha,k}(\lambda_\alpha,\nu_k)$ which acts nontrivially in the direct
product of two vector spaces $\mathbb{C}^2$: in the `horizontal' space
$\mathcal{H}_k=\mathbb{C}^2$ (associated with the $k$th horizontal line) and in the
`vertical' space $\mathcal{V}_\alpha=\mathbb{C}^2$ (associated with the
$\alpha$th vertical line). Referring to the scattering matrix picture,
the arrow states on the top and right edges of the
vertex can be regarded as `in' indices of the $L$-operator while those on
the bottom and left edges as `out' ones, that gives
\begin{equation}\label{Lop}
L_{\alpha, k}(\lambda_\alpha,\nu_k)=
a_{\alpha, k}\frac{1+\tau_\alpha^z \sigma_k^z}{2}
+b_{\alpha, k}\frac{1-\tau_\alpha^z \sigma_k^z}{2}
+ c_{\alpha, k}(\tau_\alpha^{-}\sigma_k^{+}+\tau_\alpha^{+}\sigma_k^{-}).
\end{equation}
Here $\tau_\alpha^l$ and $\sigma_k^l$ ($l=+,-,z$) denote operators
acting as Pauli matrices in $\mathcal{V}_\alpha$ and $\mathcal{H}_k$, respectively,
and identically elsewhere.

An ordered product of $L$-operators along a vertical (or horizontal) line of
the lattice corresponds in QISM to a monodromy matrix.
To construct, e.g., the vertical line monodromy matrix, it is
useful to think of the $L$-operator as a $2\times 2$ matrix acting
in space $\mathcal{V}_\alpha$, with the operator entries acting in
the space $\mathcal{H}_{1,\dots,N}=\ot_{k=1}^N\mathcal{H}_k$, i.e.,
\begin{equation}\label{L-op}
L_{\alpha,k}(\lambda,\nu)=
\begin{pmatrix}
a(\lambda,\nu)\dfrac{1+\sigma_k^z}{2}+b(\lambda,\nu)\dfrac{1-\sigma_k^z}{2} &  c\cdot\sigma_k^-\\
c\cdot\sigma_k^+ & b(\lambda,\nu)\dfrac{1+\sigma_k^z}{2}+a(\lambda,\nu)\dfrac{1-\sigma_k^z}{2}
\end{pmatrix}_{[\mathcal{V}_\alpha]}.
\end{equation}
Here the subscript indicates that this is a matrix in $\mathcal{V}_\alpha$.
The ordered product along the $\alpha$th vertical line is the `vertical'
monodromy matrix:
\begin{align}%\label{}
T_\alpha^\mathrm{V}(\lambda_\alpha)
& =
L_{\alpha,N}(\lambda_\alpha,\nu_N) \cdots
L_{\alpha,2}(\lambda_\alpha,\nu_2) L_{\alpha,1}(\lambda_\alpha,\nu_1)
\notag\\ &
=\begin{pmatrix}
A_{1,\dots,N}^\mathrm{V}(\lambda_\alpha)& B_{1,\dots,N}^\mathrm{V}(\lambda_\alpha) \\[4pt]
C_{1,\dots,N}^\mathrm{V}(\lambda_\alpha)& D_{1,\dots,N}^\mathrm{V}(\lambda_\alpha)
\end{pmatrix}_{[\mathcal{V}_\alpha]}.
\end{align}
The operators
$A_{1,\dots,N}^\mathrm{V}(\lambda)=A_{1,\dots,N}^\mathrm{V}(\lambda;\nu_1,\dots,\nu_N)$,
etc, act in $\mathcal{H}_{1,\dots,N}$ and they are independent of $\alpha$.
Each of these operators corresponds to a vertical line of the lattice, with
the top and bottom vertical arrows fixed.

Similarly, one can consider the `horizontal' monodromy matrices,
\begin{align}%\label{}
T_k^\mathrm{H}(\nu_k)
& =
L_{N,k}(\lambda_N,\nu_k) \cdots
L_{2,k}(\lambda_2,\nu_k) L_{1,k}(\lambda_1,\nu_k)
\notag\\ &
=\begin{pmatrix}
A_{1,\dots,N}^\mathrm{H}(\nu_k)& B_{1,\dots,N}^\mathrm{H}(\nu_k) \\[4pt]
C_{1,\dots,N}^\mathrm{H}(\nu_k)& D_{1,\dots,N}^\mathrm{H}(\nu_k)
\end{pmatrix}_{[\mathcal{H}_k]},
\end{align}
where operators
$A_{1,\dots,N}^\mathrm{H}(\nu)=A_{1,\dots,N}^\mathrm{H}(\nu;\lambda_1,\dots,\lambda_N)$,
etc, act in $\mathcal{V}_{1,\dots,N}:=\ot_{\alpha=1}^N\mathcal{V}_\alpha$.
Each of these operators correspond to a horizontal line of the lattice, with
the rightmost and leftmost horizontal arrows fixed.

The importance of the monodromy matrix operator entries is that they
obey a quadratic algebra, called the algebra of monodromy matrix or
Yang-Baxter algebra \cites{KBI-93}.
The algebra involves in total 16 commutation relations,
and in the following we will need some of these commutation relations, namely
\begin{equation} \label{BBCC}
B(\lambda)\, B(\lambda')=B(\lambda')\, B(\lambda),\qquad
C(\lambda)\, C(\lambda')=C(\lambda')\, C(\lambda),
\end{equation}
and
\begin{equation}\label{AB}
A(\lambda)\, B(\lambda')=
f(\lambda,\lambda')\, B(\lambda')\, A(\lambda)
+g(\lambda',\lambda)\, B(\lambda)\, A(\lambda'),
\end{equation}
where $A(\lambda)=A_{1,\dots,N}^\mathrm{V}(\lambda)$, etc, and
functions $f(\lambda',\lambda)$ and $g(\lambda',\lambda)$ are
\begin{equation}\label{fg}
f(\lambda',\lambda)=
\frac{\sin(\lambda-\lambda'+2\eta)}{\sin(\lambda-\lambda')},\qquad
g(\lambda',\lambda)=
\frac{\sin 2\eta}{\sin(\lambda-\lambda')}.
\end{equation}
Exactly the same relations are also valid for $A(\nu)=A_{1,\dots,N}^\mathrm{H}(\nu)$, etc
(after replacing $\lambda\mapsto\nu$ and $\lambda'\mapsto\nu'$ in \eqref{BBCC} and \eqref{AB}).
For the full list of commutation relations, as well as for details of their
derivation, we refer to chapter VIII of book \cite{KBI-93}.

Now we are ready to formulate the model in the framework of QISM.
Denoting by $\ket{\uparrow_k^\mathrm{V}}$ and $\ket{\downarrow_k^{\mathrm{V}}}$ the
basis vectors of space $\mathcal{H}_k$, let us introduce states
\begin{equation}\label{UpDown}
\ket{\Uparrow_{1,\dots,N}^\mathrm{V}}:=\ot_{k=1}^{N} \ket{\uparrow_k^\mathrm{V}},\qquad
\ket{\Downarrow_{1,\dots,N}^\mathrm{V}}:=\ot_{k=1}^N \ket{\downarrow_k^\mathrm{V}}.
\end{equation}
These states are the `all spins up' and `all spins down' states in the space
$\mathcal{H}_{1,\dots,N}$, respectively. Taking into account that the domain wall
boundary conditions select for $\alpha$th vertical line the operator
$B_{1,\dots,N}^\mathrm{V}(\lambda_\alpha)$, we can write the partition
function as the matrix element:
\begin{equation}\label{ZBBB}
Z_N(\lambda_1,\dots,\lambda_N;\nu_1,\dots,\nu_N)= \bra{\Downarrow_{1,\dots,N}^\mathrm{V}}
\prod_{\alpha=1}^N B_{1,\dots,N}^\mathrm{V}(\lambda_\alpha) \ket{\Uparrow_{1,\dots,N}^\mathrm{V}}.
\end{equation}
We also recall that
$B_{1,\dots,N}^\mathrm{V}(\lambda)=B_{1,\dots,N}^\mathrm{V}(\lambda;\nu_1,\dots,\nu_N)$.

Essentially in the same way, one can construct the partition function considering
operators associated with the horizontal lines. Denoting by
$\ket{\uparrow_\alpha^\mathrm{H}}$ and $\ket{\downarrow_\alpha^\mathrm{H}}$
the basis vectors of $\mathcal{V}_\alpha$, we can introduce states
\begin{equation}
\ket{\Uparrow_{1,\dots,N}^\mathrm{H}}:=\ot_{\alpha=1}^{N} \ket{\uparrow_\alpha^\mathrm{H}},\qquad
\ket{\Downarrow_{1,\dots,N}^\mathrm{H}}:=\ot_{\alpha=1}^N \ket{\downarrow_\alpha^\mathrm{H}},
\end{equation}
which are the `all spins up' and `all spins down' states of space $\mathcal{V}_{1,\dots,N}$.
The partition function reads:
\begin{equation}\label{ZCCC}
Z_N(\lambda_1,\dots,\lambda_N;\nu_1,\dots,\nu_N)= \bra{\Uparrow_{1,\dots,N}^\mathrm{H}}
\prod_{k=1}^N C_{1,\dots,N}^\mathrm{H}(\nu_k) \ket{\Downarrow_{1,\dots,N}^\mathrm{H}},
\end{equation}
and we recall that
$C_{1,\dots,N}^\mathrm{H}(\nu)=C_{1,\dots,N}^\mathrm{H}(\nu;\lambda_1,\dots,\lambda_N)$.

The partition function is known to be given by Izergin-Korepin determinant formula
(see \cites{K-82,I-87,ICK-92})
\begin{equation}\label{ZN}
Z_N=
\frac{\prod_{\alpha=1}^N \prod_{k=1}^N
a(\lambda_\alpha,\nu_k)b(\lambda_\alpha,\nu_k)}{
\prod_{1\leq\alpha<\beta\leq N}d(\lambda_\beta,\lambda_\alpha)
\prod_{1\leq j<k\leq N}d(\nu_j,\nu_k)}\,
\det_{1\leq\alpha,k\leq} \{\varphi(\lambda_\alpha,\nu_k)\}
\end{equation}
where $d(\lambda,\lambda'):=\sin(\lambda-\lambda')$ and
\begin{equation}\label{matT}
\varphi(\lambda,\nu)=\frac{c}{a(\lambda,\nu) b(\lambda,\nu)},
\end{equation}
while $a(\lambda,\nu)$, $b(\lambda,\nu)$ and $c$ are defined
in \eqref{abc}.
For the original proof of  \eqref{ZN} see \cite{ICK-92};
an alternative derivation of this formula can be found in \cites{BPZ-02,CP-07b}.

In the homogenous limit, i.e., when $\lambda_1=\cdots=\lambda_N=\lambda$ and
$\nu_1=\dots=\nu_N=0$, expression \eqref{ZN} becomes
\begin{equation}\label{ZNhom}
Z_N(\lambda,\dots,\lambda;0,\dots,0)=\frac{[\sin(\lambda-\eta)\sin(\lambda+\eta)]^{N^2}}
{\prod_{n=1}^{N-1}(n!)^2}\, \det_{1\leq\alpha,k\leq N}\{\partial_{\lambda}^{\alpha+k-2}
\varphi(\lambda)\}
\end{equation}
where $\varphi(\lambda):=\varphi(\lambda,0)$.
Below we often use simplified notations for the homogeneous model quantities, e.g., writing
$Z_N$ for $Z_{N}(\lambda,\dots,\lambda;0,\dots,0)$, and so on.

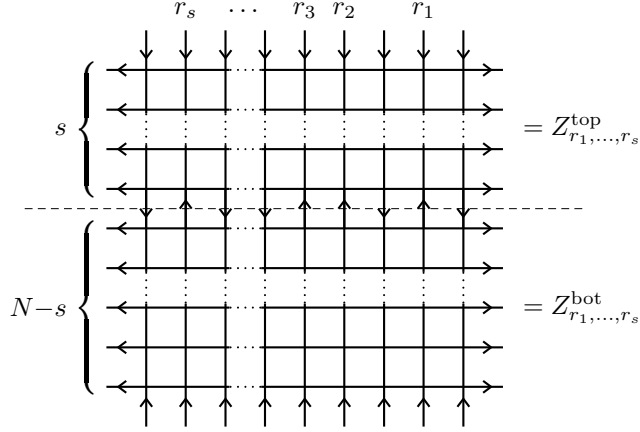
\begin{figure}
\centering
\psset{unit=15pt,dotsep=2pt}
\newcommand{\arr}{\lput{:U}{\begin{pspicture}(0,0)
\psline(0,.15)(.2,0)(0,-.15) \end{pspicture}}}

%\begin{document}
\begin{pspicture}(-2.5,0)(13.5,11)
%\psframe(-2.5,0)(13.5,11)
\multirput(1,0)(1,0){9}{\pcline(0,0)(0,1)\arr \pcline(0,10)(0,9)\arr}
\multirput(0,1)(0,1){9}{\pcline(1,0)(0,0)\arr \pcline(9,0)(10,0)\arr}
\multirput(1,1)(0,1){9}{\psline(0,0)(2.1,0)\psline[linestyle=dotted](2.1,0)(2.9,0)\psline(2.9,0)(8,0)}
\multirput(1,1)(1,0){9}{\psline(0,0)(0,2.1)\psline[linestyle=dotted](0,2.1)(0,2.9)\psline(0,2.9)(0,6.1)
    \psline[linestyle=dotted](0,6.1)(0,6.9)\psline(0,6.9)(0,8)}
\pcline(2,5)(2,6)\arr \pcline(5,5)(5,6)\arr \pcline(6,5)(6,6)\arr \pcline(8,5)(8,6)\arr
\rput(2,10.5){$r_s$}
\rput(3.5,10.5){$\dots$}
\rput(5,10.5){$r_3$}
\rput(6,10.5){$r_2$}
\rput(8,10.5){$r_1$}
\rput{-90}(-.5,7.5){$\underbrace{\hspace{1.8cm}}$}\rput[r](-1,7.5){$s$}
\rput{-90}(-.5,3){$\underbrace{\hspace{2.3cm}}$}\rput[r](-1,3){$N{-}s$}
\psline[linestyle=dashed,dash=3pt 2pt,linewidth=.02](-2.06,5.5)(12,5.5)
\rput(12,7.5){$=Z^\text{top}_{r_1,\dots,r_s}$}
\rput(12,3){$=Z^\text{bot}_{r_1,\dots,r_s}$}
\pcline(1,6)(1,5)\arr \pcline(3,6)(3,5)\arr \pcline(4,6)(4,5)\arr \pcline(7,6)(7,5)\arr \pcline(9,6)(9,5)\arr

%\rput(5,-.5){$\underbrace{\hspace{4.4cm}}$}\rput[b](5,-1.5){$N$}
%
%
%\rput(12,2){
%\multirput(1,0)(1,0){9}{\pcline(0,10)(0,9)\arr}
%\multirput(0,6)(0,1){4}{\pcline(1,0)(0,0)\arr \pcline(9,0)(10,0)\arr}
%\multirput(1,6)(0,1){4}{\psline(0,0)(2.1,0)\psline[linestyle=dotted](2.1,0)(2.9,0)\psline(2.9,0)(8,0)}
%\multirput(1,6)(1,0){9}{\psline(0,0)(0,1.1)\psline[linestyle=dotted](0,1.1)(0,1.9)\psline(0,1.9)(0,3)}
%\pcline(2,5)(2,6)\arr \pcline(5,5)(5,6)\arr \pcline(6,5)(6,6)\arr \pcline(8,5)(8,6)\arr
%\pcline(1,6)(1,5)\arr \pcline(3,6)(3,5)\arr \pcline(4,6)(4,5)\arr \pcline(7,6)(7,5)\arr \pcline(9,6)(9,5)\arr
%\rput(2,11){$r_s$}
%\rput(3.5,11){$\dots$}
%\rput(5,11){$r_3$}
%\rput(6,11){$r_2$}
%\rput(8,11){$r_1$}
%\rput(12,7.5){$=Z^\text{top}_{r_1,\dots,r_s}$}
%}
%
%\rput(12,-2){
%\multirput(1,0)(1,0){9}{\pcline(0,0)(0,1)\arr\arr}
%\multirput(0,1)(0,1){5}{\pcline(1,0)(0,0)\arr \pcline(9,0)(10,0)\arr}
%\multirput(1,1)(0,1){5}{\psline(0,0)(2.1,0)\psline[linestyle=dotted](2.1,0)(2.9,0)\psline(2.9,0)(8,0)}
%\multirput(1,1)(1,0){9}{\psline(0,0)(0,2.1)\psline[linestyle=dotted](0,2.1)(0,2.9)\psline(0,2.9)(0,4)}
%\pcline(2,5)(2,6)\arr \pcline(5,5)(5,6)\arr \pcline(6,5)(6,6)\arr \pcline(8,5)(8,6)\arr
%\pcline(1,6)(1,5)\arr \pcline(3,6)(3,5)\arr \pcline(4,6)(4,5)\arr \pcline(7,6)(7,5)\arr \pcline(9,6)(9,5)\arr
%\rput(2,7){$r_s$}
%\rput(3.5,7){$\dots$}
%\rput(5,7){$r_3$}
%\rput(6,7){$r_2$}
%\rput(8,7){$r_1$}
%\rput(12,3){$=Z^\text{bot}_{r_1,\dots,r_s}$}
%}

\end{pspicture}
%\end{document}
\caption{Row configuration with $s$ up arrows between $s$th
and $(s+1)$th horizontal lines from the top,
and the resulting partition functions for the top and bottom portions
of the original lattice.}
\label{fig.corr}
\end{figure}

Let us now turn to the row configuration probability. To define this quantity
it is useful to mention first that in the six-vertex model with domain wall boundary
conditions all configurations are such that on the $s$th row (i.e., on the
$N$ vertical edges between the $s$th and the $(s+1)$th horizontal lines,
counted from the top, in our conventions) there are exactly $s$ arrows
pointing up. It is therefore natural to study the probability of observing a
given configuration of arrows on a given row, or row configuration
probability, for short. Namely, we denote by $H_{N,s}^{(r_1,\dots,r_s)}$
the probability that the $s$ up-arrows of the $s$th row are
exactly at the positions $r_1,\dots,r_s$ (counted from the right), see
figure~\ref{fig.corr}.

Since the row configuration probability describes generic configurations of the model,
it can be used as a building block to compute other correlation functions. For
example, by properly summing over positions of up-arrows, one can recover the
so-called the emptiness formation probability studied in \cite{CP-07b}. This
connection is discussed in section 5.

To compute the row configuration probability,
we separate the original $N\times N$ lattice
into two smaller lattices: an upper lattice, with $s$ horizontal and $N$
vertical lines, and a lower lattice, with $N-s$ horizontal and $N$ vertical
lines. We shall denote $\Ztop{r_1,\dots,r_s}$  and $\Zbot{r_1,\dots,r_s}$ the
partition functions of  the  six-vertex model on the upper and lower
sublattices, respectively (see figure~\ref{fig.corr}).
The row configuration probability is essentially given
as a product of the partition functions
of the six-vertex model on these two smaller lattices,
\begin{equation}\label{defHNs}
H_{N,s}^{(r_1,\dots,r_s)}=\frac{\Ztop{r_1,\dots,r_s}
\, \Zbot{r_1,\dots,r_s}}{Z_N}.
\end{equation}
Our main goal in the present paper is therefore the derivation of some
useful representations for the partition functions $\Ztop{r_1,\dots,r_s}$ and
$\Zbot{r_1,\dots,r_s}$. Specifically, we provide multiple ($s$-fold) integral
representations for these quantities.

In terms of QISM objects, the partition functions on the upper, $s\times N$,
sublattice can be written similarly to representation \eqref{ZCCC}, as
follows:
\begin{equation}\label{defZtop}
\Ztop{r_1,\dots,r_s}=\bra{\Downarrow_{1,\dots,N}^\mathrm{H}}
\tau_{r_s}^{-}\cdots \tau_{r_2}^{-}\tau_{r_1}^{-}
\prod_{k=1}^s C_{1,\dots,N}^\mathrm{H}(\nu_k)
\ket{\Downarrow_{1,\dots,N}^\mathrm{H}},
\end{equation}
where $\tau_j^{-}$ ($j=1,\dots,N$), as above, denote Pauli matrices acting in
spaces $\mathcal{V}_j$.

To write the partition function of the lower, $(N-s)\times N$, sublattice as
a matrix element, let us define vectors:
\begin{equation}
\ket{\Uparrow_{s+1,\dots,N}^\mathrm{V}}=\ot_{k=s+1}^N \ket{\uparrow_k^\mathrm{V}},
\qquad
\ket{\Downarrow_{s+1,\dots,N}^\mathrm{V}}=\ot_{k=s+1}^N \ket{\downarrow_k^\mathrm{V}}.
\end{equation}
These are `all spins up' and `all spin down' states of the space
$\mathcal{H}_{s+1,\dots,N}$. Correspondingly, let us consider the matrix
elements of the `truncated' vertical monodromy matrix given as the product
$L_{\alpha,N}(\lambda_\alpha,\nu_N) \cdots
L_{\alpha,s+1}(\lambda_\alpha,\nu_{s+1})$. These matrix elements are
operators $A_{s+1,\dots,N}^\mathrm{V}(\lambda_\alpha)
=A_{s+1,\dots,N}^\mathrm{V}(\lambda_\alpha;\nu_{s+1},\dots,\nu_N)$,
etc, acting in $\mathcal{H}_{s+1,\dots,N}$. The partition function
$\Zbot{r_1,\dots,r_s}$ can be written as
\begin{multline}\label{defZbot}
\Zbot{r_1,\dots,r_s}=
\bra{\Downarrow_{s+1,\dots,N}^\mathrm{V}}
\prod_{\alpha=r_s+1}^{N}
B(\lambda_{\alpha})
\cdot
A(\lambda_{r_s})\cdot
\prod_{\alpha=r_{s-1}+1}^{r_s-1}
B(\lambda_{\alpha})
\\ \times \cdots \times
A(\lambda_{r_2})
\prod_{\alpha=r_1+1}^{r_2-1}
B(\lambda_{\alpha})\cdot
A(\lambda_{r_1})\cdot
\prod_{\alpha=1}^{r_1-1}
B(\lambda_{\alpha})
\ket{\Uparrow_{s+1,\dots,N}^\mathrm{V}},
\end{multline}
where $A(\lambda):=A_{s+1,\dots,N}^\mathrm{V}(\lambda)$ and
$B(\lambda):=B_{s+1,\dots,N}^\mathrm{V}(\lambda)$.

%%%%%%%%%%%%%%%%%%%%%%%%%%%%%%%%%%%%%%%%%%%%%%%%%%%%%%%%%%%%%%%%%%%%
\section{Calculation of $\Ztop{r_1,\dots,r_s}$}

The matrix element in \eqref{defZtop} can be formally evaluated
(as a function of $\lambda_1,\dots,\lambda_N$ and $\nu_1,\dots,\nu_s$) using the
equivalence of the algebraic and coordinate versions of
the Bethe Ansatz. This equivalence was first explicitly proved, as a side result,
in \cite{IKR-87} (see appendix D of that paper); see also book \cite{KBI-93}, Chapter VII.

For simplicity, we start directly from the case where parameters $\lambda_1,\dots,\lambda_N$
are already taken to the same value $\lambda$, but
the parameters $\nu_1,\dots,\nu_s$ are left arbitrary (and not equal to each other).
Equation (D.4) of reference \cite{IKR-87} in such a case implies
\begin{align}\label{oldrep}
\Ztop{r_1,\dots,r_s}
&
=c^s \prod_{k=1}^s \left[a(\lambda,\nu_k)\right]^{N-1}
\prod_{1\leq j<k\leq s}\frac{1}{t_k-t_j}
\notag\\ &\quad\times
\sum_{P\in\Omega_s} (-1)^{[P]}
\prod_{j=1}^s t_{P_j}^{r_j-1}\prod_{1\leq j < k\leq s}
(t_{P_j}t_{P_k}-2\Delta t_{P_j}+1),
\end{align}
where
\begin{equation}
t_k:=\frac{b(\lambda,\nu_k)}{a(\lambda,\nu_k)},
\end{equation}
and the sum is taken over elements of the symmetric group
$\Omega_s$, i.e., permutations $P:1,\dots, s\mapsto P_1,\dots,P_s$,
with $[P]$ denoting the parity of $P$. Clearly, the expression standing
in the second line in \eqref{oldrep} is exactly the $s$-particle coordinate
Bethe Ansatz trial wave-function.

To study the homogeneous limit of \eqref{oldrep} in the remaining set of
parameters, we first transform slightly this expression. Let
us set $t_k=t+x_k$ ($k=1,\dots,s$) where $t$ is an arbitrary parameter and the new
parameters $x_1,\dots,x_s$ are all different. Using the fact that for a function
$f(x)$, regular near point $x=t$, one can always write
$f(t+x)=\exp(x\partial_z) f(t+z)|_{z=0}$, we bring \eqref{oldrep} to the form
\begin{align}\label{oldrep_det}
\Ztop{r_1,\dots,r_s}
&
=c^s \prod_{k=1}^s \left[a(\lambda,\nu_k)\right]^{N-1}
\prod_{1\leq j<k\leq s}\frac{1}{x_k-x_j}
\det_{1\leq j,k\leq s}\left\{\exp\big(x_j\partial_{z_k}\big)\right\}
\notag\\  &\quad\times
\prod_{j=1}^s (t+z_j)^{r_j-1}
\prod_{1\leq j<k\leq s}\left[(t+z_j)(t+z_k) -2\Delta(t+z_j) +1\right]\bigg|_{z_1=\dots=z_s=0}
\end{align}
which simply represents an equivalent way to write \eqref{oldrep}.

Let us now consider the homogeneous limit in the parameters $\nu_1,\dots,\nu_s$.
Since $t$ is arbitrary, we can perform this limit such that
$t_1=\dots=t_k=t$ in the limit, and put $t=b/a$, where $a$ and $b$ are
the homogeneous model weights, see \eqref{abc}. We thus have to consider \eqref{oldrep}
at $x_1=\dots=x_s=0$. The limit in \eqref{oldrep_det} can be done using the relation
\begin{equation}
\frac{\det_{1\leq j,k\leq s}\{\exp(x_j\partial_{z_k})\}}{\prod_{1\leq j<k\leq s}(x_k-x_j)}\Bigg|_{x_1=\dots=x_s=0}
=\det_{1\leq j,k\leq s}\left\{\frac{1}{(j-1)!}\partial_{z_k}^{j-1}\right\}.
\end{equation}
Reexpressing the values of derivatives at $z_1=\dots=z_s=0$ as residues, we
obtain a multiple integral representation
\begin{align}\label{oldrep_hom}
\Ztop{r_1,\dots,r_s}
&
=c^s a^{s(N-1)}\oint_{C_0}\dots\oint_{C_0}
\prod_{j=1}^s (t+z_j)^{r_j-1}\det_{1\leq j,k\leq s}\left\{z_k^{-j}\right\}
\notag\\  &\quad\times
\prod_{1\leq j<k\leq s}\left[(t+z_j)(t+z_k) -2\Delta(t+z_j) +1\right]
\frac{\rmd^s z}{(2\rmi \pi)^s}.
\end{align}
Here $C_0$ is a small, simple, closed, positively-oriented contour enclosing point
$z=0$. Evaluating the Vandermonde determinant and making the change
$z_k\mapsto w_k=(z_k+t)/t$, we finally obtain:
\begin{align}\label{oldrep_mir}
\Ztop{r_1,\dots,r_s}
&
=c^s a^{s(N-1)}
\prod_{j=1}^s t^{r_j-j}
\oint_{C_1}^{} \cdots \oint_{C_1}^{} \prod_{j=1}^s \frac{w_j^{r_j-1}}{(w_j-1)^s}
\notag\\ &\quad\times
\prod_{1\leq j<k\leq s} \left[(w_j-w_k)(t^2 w_j w_k -2\Delta t w_j +1)\right]
\frac{\rmd^s w}{(2\pi\rmi)^s}.
\end{align}
Here $C_1$ denotes a small, simple, closed, positively-oriented contour enclosing point $z=1$.

Formulae \eqref{oldrep} and \eqref{oldrep_mir} can also been derived by other
methods (i.e., without using the equivalence of the algebraic and coordinate
Bethe Ansatz), e.g., starting with
vertical monodromy matrix formulation of $\Ztop{r_1,\dots,r_s}$,
analogous
to \eqref{defZbot} for $\Zbot{r_1,\dots,r_s}$,
and next using the technique of paper \cite{KMT-99}
to evaluate the matrix element\footnote{We are indebted to P. Zinn-Justin
for explaining us this alternative derivation.}.
We also mention that
formula \eqref{oldrep_det}, in a different form and
for special values of $t$ and $\Delta$, has been found in the context of
enumerative combinatorics \cite{F-06}.

%%%%%%%%%%%%%%%%%%%%%%%%%%%%%%%%%%%%%%%%%%%%%%%%%%%%%%%%%%%%%%%%%%%%
\section{Calculation of $\Zbot{r_1,\dots,r_s}$}

Taking into account commutativity of $B$-operators, see \eqref{BBCC},
and using relation \eqref{AB}, we
can obtain, in the usual spirit of the algebraic Bethe Ansatz calculation
(for details see, e.g., \cites{KBI-93}), the relation:
\begin{equation}\label{derivationAB}
A(\lambda_r)\prod_{\beta=1}^{r-1} B(\lambda_\beta)=
\sum_{\alpha=1}^r
\frac{g(\lambda_\alpha,\lambda_r)}{f(\lambda_\alpha,\lambda_r)}
\prod_{\substack{\beta=1\\ \beta\ne\alpha}}^{r} f(\lambda_\alpha,\lambda_\beta)
\prod_{\substack{\beta=1\\ \beta\ne\alpha}}^{r} B(\lambda_\beta) A(\lambda_\alpha).
\end{equation}
Using this commutation relation and taking into account that
\begin{equation}\label{Avac}
A_{s+1,\dots,N}^\mathrm{V}(\lambda)\ket{\Uparrow_{s+1,\dots,N}^\mathrm{V}}
=\prod_{k=s+1}^N a(\lambda,\nu_k)\ket{\Uparrow_{s+1,\dots,N}^\mathrm{V}},
\end{equation}
and also using \eqref{ZBBB}, we obtain
\begin{multline}\label{derivation1}
\Zbot{r_1,\dots,r_s}=\sum_{\alpha_1=1}^{r_1}
\sum_{\substack{\alpha_2=1\\ \alpha_2\ne\alpha_1}}^{r_2}
\cdots
\sum_{\substack{\alpha_s=1\\ \alpha_s\ne\alpha_1,\,\dots,\alpha_{s-1}}}^{r_s}
\prod_{j=1}^s\prod_{k=s+1}^N  a(\lambda_{\alpha_j},\nu_k)
\prod_{j=1}^s\frac{g(\lambda_{\alpha_j},\lambda_{r_j})}
{f(\lambda_{\alpha_j},\lambda_{r_j})}
\\
\times
\prod_{\substack{\beta_1=1\\ \beta_1 \ne\alpha_1}}^{r_1}
f(\lambda_{\alpha_1},\lambda_{\beta_1})
\prod_{\substack{\beta_2=1\\ \beta_2 \ne\alpha_1,\alpha_2}}^{r_2}
f(\lambda_{\alpha_2},\lambda_{\beta_2})
\ \cdots
\prod_{\substack{\beta_s=1\\ \beta_s \ne\alpha_1,\dots,\alpha_s}}^{r_s}
f(\lambda_{\alpha_s},\lambda_{\beta_s})
\\
\times
Z_{N-s}(\lambda_{\bar\alpha_1},\dots,\lambda_{\bar\alpha_{N-s}};\nu_{s+1},\dots,\nu_N),
\end{multline}
where $\{\bar\alpha_1,\dots,\bar\alpha_{N-s}\}:=\{1,\dots,N\}\backslash\{\alpha_1,\dots,\alpha_s\}$.

To proceed further, it is convenient to introduce function:
\begin{equation}\label{vrs}
v_{r}(\lambda)=\frac{\prod_{\alpha=r+1}^N d(\lambda_\alpha,\lambda)
\prod_{\alpha=1}^{r-1} e(\lambda_\alpha,\lambda)}{\prod_{k=s+1}^N b(\lambda,\nu_k)},
\end{equation}
where $d(\lambda,\lambda'):=\sin(\lambda-\lambda')$ and $e(\lambda,\lambda'):=\sin(\lambda-\lambda'+2\eta)$.
Expressing functions  $f(\lambda,\lambda')$ and  $g(\lambda,\lambda')$
appearing in \eqref{derivation1} in terms of functions
$d(\lambda,\lambda')$ and $e(\lambda,\lambda')$ and
substituting the  Izergin-Korepin determinant formula, see \eqref{ZN}, for
the partition function standing in \eqref{derivation1},
we arrive at the expression:
\begin{multline}\label{derivation2}
\Zbot{r_1,\dots,r_s}=
\frac{
\prod_{\alpha=1}^N\prod_{k=s+1}^N
a(\lambda_{\alpha},\nu_k)b(\lambda_\alpha,\nu_k)}
{\prod_{1\leq\alpha<\beta\leq N}d(\lambda_\beta,\lambda_\alpha)
\prod_{s+1\leq j<k\leq N} d(\nu_j,\nu_k)}
\\
\times\sum_{\alpha_1}^{r_1}
\sum_{\substack{\alpha_2=1\\ \alpha_2\ne\alpha_1}}^{r_2}
\cdots
\sum_{\substack{\alpha_s=1\\ \alpha_s\ne\alpha_1,\dots,\alpha_{s-1}}}^{r_s}
(-1)^{\sum_{j=1}^s(\alpha_j-1)-\sum_{1\leq j<k\leq s} \chi(\alpha_k,\alpha_j)}
\\
\times
\prod_{j=1}^s v_{r_j}(\lambda_{\alpha_j})
\prod_{1\leq j<k\leq s}\frac{1}{e(\lambda_{\alpha_j},\lambda_{\alpha_k})}
\det_{1\leq j,k \leq N-s}\{\varphi(\lambda_{\bar\alpha_j},\nu_{s+k})\}.
\end{multline}
Here $\chi(\beta,\alpha)=1$ if $\beta>\alpha$, and $\chi(\beta,\alpha)=0$ otherwise.

Clearly, the multiple sum in \eqref{derivation2} reminds the Laplace
expansion of some $N\times N$ determinant. This is also in agreement with the
fact that since $v_{r}(\lambda_\alpha)=0$ ($\alpha=r+1,\dots,N$) all
summations in \eqref{derivation2} can be extended till the value $N$. To
write down such a determinant formula, let us set
$\lambda_\alpha=\lambda+\xi_\alpha$ ($\alpha=1,\dots,N$), where $\lambda$ is
some arbitrary parameter, and parameters $\xi_1,\dots,\xi_N$ are all different. Using
again the fact that for a function $f(\xi)$, regular near point
$\xi=\lambda$, we can write $f(\lambda+\xi)=\exp(\xi\partial_\eps)
f(\lambda+\eps)|_{\eps=0}$, we can bring \eqref{derivation2} to the form
\begin{multline}\label{Z2det}
\Zbot{r_1,\dots,r_s}=
\frac{
\prod_{\alpha=1}^N\prod_{k=s+1}^N
a(\lambda_{\alpha},\nu_k)b(\lambda_\alpha,\nu_k)}
{\prod_{1\leq\alpha<\beta\leq N}d(\lambda_\beta,\lambda_\alpha)
\prod_{s+1\leq j<k\leq N} d(\nu_j,\nu_k)}
\\ \times
\begin{vmatrix}
\exp(\xi_1\partial_{\eps_1}) & \dots & \exp(\xi_1\partial_{\eps_s})
& \varphi(\lambda_1,\nu_{s+1}) & \dots & \varphi(\lambda_1,\nu_{N})
\\
\exp(\xi_2\partial_{\eps_1}) & \dots & \exp(\xi_2\partial_{\eps_s})
& \varphi(\lambda_2,\nu_{s+1}) & \dots & \varphi(\lambda_2,\nu_{N})
\\
\hdotsfor{6}
\\
\exp(\xi_N\partial_{\eps_1}) & \dots & \exp(\xi_N\partial_{\eps_s})
& \varphi(\lambda_N,\nu_{s+1}) & \dots & \varphi(\lambda_N,\nu_{N})
\end{vmatrix}
\\ \times\prod_{j=1}^s v_{r_j}(\lambda+\eps_j)
\prod_{1\leq j<k\leq s}^{}
\frac{1}
{e(\lambda+\eps_j,\lambda+\eps_k)}\Bigg|_{\eps_1=\ldots=\eps_s=0}.
\end{multline}
We stress that this expression is valid for the inhomogeneous model.

Let us now perform the homogeneous limit. We
regard $\lambda$ as the parameter of the weights of the homogeneous model, so that
parameters $\xi_1,\dots,\xi_N$ and $\nu_{s+1},\dots,\nu_N$ are sent to zero in the limit.
The procedure can be done along the lines
of \cite{ICK-92} and it is explained in full detail in \cite{CP-07b}.
As a result, we obtain the expression
\begin{align}\label{homZ2}
\Zbot{r_1,\dots,r_s}
&
=\frac{(ab)^{N(N-s)}}{\prod_{j=1}^{N-s-1}j!\prod_{k=1}^{N-1}k!}
\begin{vmatrix}
\varphi(\lambda)& \dots & \partial_\lambda^{N-s-1}\varphi(\lambda)& 1& \dots &  1
\\
\partial_\lambda\varphi(\lambda) & \dots & \partial_\lambda^{N-s}\varphi(\lambda)
& \partial_{\eps_1} & \dots & \partial_{\eps_s}
\\ \hdotsfor{6} \\
\partial_\lambda^{N-1}\varphi(\lambda) & \dots &
\partial_\lambda^{2N-s-2}\varphi(\lambda)
& \partial_{\eps_1}^{N-1} & \dots & \partial_{\eps_s}^{N-1}
\end{vmatrix}
\notag\\ &\quad \times
\prod_{j=1}^{s}
\frac{(\sin\eps_j)^{N-r_j}[\sin(\eps_j-2\eta)]^{r_j-1}}{[\sin(\eps_j+\lambda-\eta)]^{N-s}}
\!\!\!\!\prod_{1\leq j<k \leq s}^{} \frac{1}{\sin(\eps_j-\eps_k+2\eta)}
\Bigg|_{\eps_1=\ldots=\eps_s=0},
\end{align}
where in writing the determinant we have changed the order of columns,
in comparison with \eqref{Z2det}.

In order to represent \eqref{homZ2} in terms of a multiple integral, we first
transform the $N\times N$ determinant representation \eqref{homZ2} to an
$s\times s$ one, given in terms of certain set of orthogonal polynomials.
The construction is based on the following general facts from the theory of
orthogonal polynomials (see, e.g., \cite{S-75}).
Let $\{P_n(x)\}_{n=0}^\infty$ be a set of orthogonal polynomials,
\begin{equation}\label{ortho}
\int   P_{n}(x) P_{m}(x) \mu(x)\,\rmd x
=h_{n} \delta_{nm} ,
\end{equation}
where the integration domain is assumed over the real axis and we
choose normalisation such that
$P_n(x)=x^n + \dots$, and let $c_n$ denote the $n$th
moment of the weight $\mu(x)$,
\begin{equation}%\label{}
c_n =\int x^n \mu(x)\, \rmd x
\qquad (n=0,1,\ldots).
\end{equation}
Then $\det_{1\leq j,k\leq N}\{c_{j+k-2}\}=h_0h_1\cdots h_{N-1}$ and, more
generally, for $s=1,\dots,N$, the following formula is valid:
\begin{multline}\label{detdet}
\begin{vmatrix}
c_0&c_1& \dots &c_{N-s-1}  & 1 & 1 & \dots & 1 \\
c_1&c_2& \dots &c_{N-s} & x_1 & x_2 &\dots &x_s \\
\hdotsfor{8}  \\
c_{N-1}& c_{N} &\dots & c_{2N-s-2} & x_1^{N-1} & x_2^{N-1} &\dots &x_s^{N-1}
\end{vmatrix}
\\
= h_0 h_1\cdots h_{N-s-1}
\det_{1\leq j,k\leq s} \{P_{N-s+j-1}(x_k)\}.
\end{multline}
In our case $c_n:=\partial_\lambda^n\varphi(\lambda)$, and the integration
measure $\mu(x)\rmd x$ can be
found through the Laplace transform for function $\varphi(\lambda)$; for
explicit expressions, see \cite{Zj-00}.

As in \cites{CP-07b}, we denote
\begin{equation}\label{Knx}
K_n(x)= \frac{n!\,\varphi^{n+1}}{h_n}\; P_n(x),
\end{equation}
where $\varphi:=\varphi(\lambda)$, and $h_n$ is as in \eqref{ortho}.
We also introduce functions
\begin{equation}\label{omegas}
\omega(\epsilon):=\frac{a}{b}\,
\frac{\sin\eps}{\sin(\eps-2\eta)},\qquad
\tilde\omega(\epsilon):=\frac{b}{a}\,
\frac{\sin\eps}{\sin(\eps+2\eta)}
\end{equation}
which, in particular, satisfy the relation
\begin{equation}%\label{}
\frac{b}{c}\,
\frac{\sin(\eps-2\eta)}{\sin(\eps+\lambda-\eta)}=\frac{1}{\omega(\eps)-1},
\end{equation}
where the short notations for the weights $a:=\sin(\lambda+\eta)$,
$b:=\sin(\lambda-\eta)$, and $c:=\sin2\eta$ are used. Taking into account
that $\varphi=c/ab$, and using the relation
\begin{equation}%\label{}
\frac{\sin(\eps_1+\lambda+\eta)\sin(\eps_2+\lambda-\eta)}
{\sin(\eps_1-\eps_2+2\eta)}
=\frac{1}{\varphi}\;
\frac{(1-\tilde\omega(\eps_1))(\omega(\eps_2)-1)}{\tilde\omega(\eps_1)\omega(\eps_2)-1},
\end{equation}
after applying \eqref{detdet} to \eqref{homZ2}, we obtain:
\begin{align}\label{orthZ2}
\Zbot{r_1,\dots,r_s}
&
=\frac{Z_N}{a^{\frac{s(2N-s+1)}{2}}
b^{\frac{s(s-3)}{2}}c^s}\left(\frac{a}{b}\right)^{r_1+\dots+r_s}
\det_{1\leq j,k\leq s} \left\{K_{N-s+j-1}\big(\partial_{\eps_k}\big)\right\}
\notag\\ &\quad\times
\prod_{j=1}^{s}\frac{[\omega(\eps_j)]^{N-r_j-s+j}
[\tilde\omega(\eps_j)]^{s-j}}{[\omega(\eps_j)-1]^{N-s}}
\prod_{1\leq j<k \leq s}^{} \frac{1}
{\tilde\omega(\eps_j)\omega(\eps_k)-1}
\Bigg|_{\eps_1=\ldots=\eps_s=0}.
\end{align}
In deriving of this formula, we have also used \eqref{ZNhom} to express
a proper factor as the partition function $Z_N$.

Now we are ready to write representation \eqref{orthZ2} as a multiple integral.
We follow the procedure developed in \cite{CP-07b}. The key relation here, valid
for an arbitrary function $f(z)$ regular at the origin, is
\begin{equation}\label{claim}
K_{N-1}(\partial_\eps)\, f(\omega(\eps))\Big|_{\eps=0}=
\frac{1}{2\pi \rmi}\oint_{C_0}^{} \frac{(z-1)^{N-1}}{z^N} h_N(z) f(z)\, \rmd z.
\end{equation}
Here $C_0$, as above, is a small, simple, closed, positively-oriented contour enclosing point
$z=0$, and $h_N(z)$ (not to be confused with  $h_n$ in \eqref{ortho}) is the
generating function for the one-point boundary correlation function,
$h_N(z)=\sum_{r=1}^{N} H_N^{(r)}z^{r-1}$, where
\begin{equation}\label{H_N}
H_N^{(r)}=K_{N-1}(\partial_\eps)\,\frac{[\omega(\eps)]^{N-r}}{[\omega(\eps)-1]^{N-1}}
\bigg|_{\epsilon=0}.
\end{equation}
This function can be viewed as the $s=1$ case of the row configuration probability,
$H_N^{(r)}:=H_{N,1}^{(r)}$. Indeed, in this case the partition function of
the upper sublattice is simply $\Ztop{r}=a^{N-r}b^{r-1} c$, while $\Zbot{r}$
can be found from \eqref{orthZ2}, thus reproducing \eqref{H_N}.

To write down the resulting multiple integral representation for
$\Zbot{r_1,\dots,r_s}$, we introduce functions
\begin{equation}\label{hNs}
h_{N,s}(z_1,\dots,z_s) =
\frac{\det_{1\leq j,k \leq s}\{z_k^{s-j} (z_k-1)^{j-1} h_{N-s+j}(z_k)\}}{
\prod_{1\leq j<k \leq s}^{} (z_k-z_j)},
\end{equation}
which can be viewed as multi-variable generalisations of $h_N(z)$ (for a
detailed discussion of its properties, see \cite{CP-07b}). Noticing that
\begin{equation}\label{tildez}
\tilde \omega(\epsilon) = \frac{t^2\omega(\eps)}{2\Delta t \omega(\eps) -1},
\end{equation}
where $t=b/a$ and $\Delta=(a^2+b^2-c^2)/2ab$,
we can readily rewrite the orthogonal
polynomial representation \eqref{orthZ2} in virtue of \eqref{claim} as follows:
\begin{multline}\label{MIRZ2}
\Zbot{r_1,\dots,r_s}=
Z_N\frac{\prod_{j=1}^{s}t^{j-r_j}}{a^{s(N-1)}c^s}
\oint_{C_0}^{} \cdots \oint_{C_0}^{}
\prod_{j=1}^s\frac{1}{z^{r_j}_j}
\prod_{1\leq j <k \leq s}\frac{z_k-z_j}{t^2 z_j z_k -2\Delta t z_j+1}
\\ \times
h_{N,s}(z_1,\dots,z_s)
\, \frac{\rmd^s z}{(2\pi \rmi)^s}.
\end{multline}
This formula is the desired representation for $\Zbot{r_1,\dots,r_s}$,
valid for the homogeneous model.

%%%%%%%%%%%%%%%%%%%%%%%%%%%%%%%%%%%%%%%%%%%%%%%%%%%%%%%%%%%%%%%%%%%%
\section{Emptiness formation probability}

An important example of correlation function which can be built from the row
configuration probability is the emptiness formation probability. As in
\cite{CP-07b}, we denote by $F_N^{(r,s)}$ the probability of observing all arrows on the first
$s$ horizontal edges (counted, as usual, from the top) located between $r$-th and
$(r+1)$-th vertical lines (counted, as usual, from the right) to be all
pointing left. Equivalently, due to both the domain wall boundary conditions
and the ice-rule, we can define it as the probability of observing
the last $N-r$ arrows between the $s$th
and $(s+1)$th horizontal lines to be all pointing down, and hence
(see also figure~\ref{fig.corr}) we have the relation:
\begin{equation}\label{efp}
F_N^{(r,s)}=
\sum_{1\leq r_1 < r_2 < \dots < r_s \leq r}
H_{N,s}^{(r_1,\dots,r_s)}.
\end{equation}
Our aim here is to address how this summation can be done for
the row configuration probability, given by \eqref{defHNs},
\eqref{oldrep_mir} and \eqref{MIRZ2}, to reproduce
the multiple integral representations
for the emptiness formation probability obtained in \cite{CP-07b}.

Let us first recall the results of paper \cite{CP-07b}. The
following two multiple integral representations have been obtained:
\begin{align}\label{efpMIR}
F_N^{(r,s)}
&
= (-1)^s\oint_{C_0}^{} \cdots \oint_{C_0}^{}
\prod_{j=1}^{s}\frac{[(t^2-2\Delta t)z_j+1]^{s-j}}{z_j^r(z_j-1)^{s-j+1}}\,
\prod_{1\leq j<k \leq s}^{} \frac{z_j-z_k}{t^2z_jz_k-2\Delta t z_j+1}
\notag\\ &\quad\times
h_{N,s}(z_1,\dots,z_s)
\,\frac{\rmd^s z}{(2\pi \rmi)^s}
\notag\\ &
=\frac{(-1)^{s}Z_s}{s! a^{s(s-1)}c^s}
\oint_{C_0}^{} \cdots \oint_{C_0}^{}
\prod_{j=1}^{s} \frac{[(t^2-2\Delta t)z_j+1]^{s-1}}{z_j^r(z_j-1)^s}\,
\prod_{\substack{j,k=1\\ j\ne k}}^{s} \frac{z_k-z_j}{t^2 z_jz_k-2\Delta t z_j +1}
\notag\\ &\quad\times
h_{N,s}(z_1,\dots,z_s)
h_{s,s}(u(z_1),\dots,u(z_s))
\,\frac{\rmd^s z}{(2\pi\rmi)^s},
\end{align}
where
\begin{equation}\label{uofz}
u(z):=-\frac{z-1}{(t^2-2\Delta t)z+1}.
\end{equation}
The two representations in \eqref{efpMIR} are related by
a symmetrization of the integrand, which is due to
the following relation
\begin{multline}\label{identity1}
\Asym_{z_1,\dots,z_s}\left[
\prod_{1\leq j < k \leq s}
\frac{[(t^2-2\Delta t)z_j+1](t^2 z_j z_k -2\Delta t z_k+1)}{(z_j-1)}
\right]
\\
=\frac{Z_s}{s!a^{s(s-1)}c^s}
\prod_{j=1}^s\frac{[(t^2-2\Delta t)z_j+1]^{s-1}}{(z_j-1)^{s-1}}
\prod_{1\leq j < k \leq s}(z_k-z_j)
\,h_{s,s}(u(z_1),\dots,u(z_s)).
\end{multline}
Here $\Asym_{z_1,\dots,z_s}f(z_1,\dots,z_s):=\frac{1}{s!}\sum_{P}(-1)^{[P]}
f(z_{P_1},\dots,z_{P_s})$, and the sum is taken over permutations $P:1,\dots,
s\mapsto P_1,\dots,P_s$, with $[P]$ denoting the parity of $P$.
For details on the proof of relation \eqref{identity1}, see \cites{CP-07b}.

Let us discuss the  representation for emptiness formation probability
obtained from \eqref{defHNs}, \eqref{oldrep_mir} and \eqref{MIRZ2},
according to relation \eqref{efp}. Direct substitution gives
\begin{align}\label{efpdoubleMIR}
F_N^{(r,s)}&=
\oint_{C_1}^{} \cdots \oint_{C_1}^{}
\frac{\rmd^s w}{(2\pi \rmi)^s}
\oint_{C_0}^{} \cdots \oint_{C_0}^{}
\prod_{j=1}^s \frac{1}{(w_j-1)^s}
\sum_{1\leq r_1 < r_2 < \dots < r_s \leq r}
\prod_{j=1}^s \frac{w_j^{r_j-1}}{z_j^{r_j}}
\notag\\ &\quad\times\!\!
\prod_{1\leq j<k\leq s}
\frac{(w_j-w_k)(t^2 w_j w_k -2\Delta t w_j +1)(z_k-z_j)}{t^2 z_j z_k -2 \Delta t z_j +1}
h_{N,s}(z_1,\dots,z_s)\frac{\rmd^s z}{(2\pi \rmi)^s}
\end{align}
so performing here the multiple sum
and integrating over a set of variables (e.g., over $w_1,\dots,w_s$)
we should reproduce, modulo symmetrization of the integrand,
the $s$-fold integral representations \eqref{efpMIR}.

To address this problem, let us first consider the evaluation of the multiple sum
in \eqref{efpdoubleMIR}. Observing that the integral over $z_j$ vanish for $r_j\leq 0$, because
in this case the integrand is regular at $z_j=0$, we
can replace the sum in \eqref{efpdoubleMIR} over values
$1\leq r_1<r_2\cdots<r_s\leq r$ with a sum over values $-\infty< r_1<r_2\cdots<r_s\leq r$.
Then, denoting $X_j=z_j/w_j$, the summation can be done using the identity
\begin{equation}
\sum_{-\infty< r_1 < r_2 < \dots < r_s \leq r}
\prod_{j=1}^s \frac{1}{X_j^{r_j}}
=\prod_{j=1}^s \frac{1}{X_j^{r-s+j}\big(1-\prod_{l=1}^j X_l\big)},
\end{equation}
which can be easily verified by expanding the  denominators in the right hand side
in Taylor series. As a result, we find that \eqref{efpdoubleMIR}
simplifies to expression:
\begin{align}\label{efpdoubleMIR2}
F_N^{(r,s)}&=
\oint_{C_1}^{} \cdots \oint_{C_1}^{}
\frac{\rmd^s w}{(2\pi \rmi)^s}
\oint_{C_0}^{} \cdots \oint_{C_0}^{}
\prod_{j=1}^s \frac{w_j^r}{(w_j-1)^s z_j^{r-s+j} (\prod_{l=1}^j w_l-\prod_{l=1}^j z_l)}
\notag\\ &\quad\times\!\!
\prod_{1\leq j<k\leq s}
\frac{(w_j-w_k)(t^2 w_j w_k -2\Delta t w_j +1)(z_k-z_j)}{t^2 z_j z_k -2 \Delta t z_j +1}
h_{N,s}(z_1,\dots,z_s)\frac{\rmd^s z}{(2\pi \rmi)^s}
\end{align}
and we are left with performing an $s$-fold integration.

We shall integrate over variables $w_1,\dots,w_s$ in \eqref{efpdoubleMIR2}.
Let us consider the equivalent integral where the integrand is symmetrized
with respect to permutations of these variables. Define function
\begin{multline}
\varPhi_s(w_1,\dots,w_s;z_1,\dots,z_s)=
\prod_{1\leq j<k\leq s}\frac{1}{w_k-w_j}
\\ \times
\Asym_{w_1,\dots,w_s}\Bigg[
\frac{\prod_{1\leq j < k \leq s}(t^2 w_j w_k -2\Delta t w_j +1)}
{\prod_{j=1}^s(\prod_{l=1}^j w_l-\prod_{l=1}^j z_l)}
\Bigg].
\end{multline}
Integration over the $w_j$'s is done with the result
\begin{multline}%\label{}
\oint_{C_1}^{} \cdots \oint_{C_1}^{}
\prod_{j=1}^s \frac{w_j^r}{(w_j-1)^s}
\prod_{1\leq j<k\leq s}
(w_j-w_k)^2\; \varPhi_s(w_1,\dots,w_s;z_1,\dots,z_s)
\frac{\rmd^s w}{(2\pi \rmi)^s}
\\
=(-1)^{s(s-1)/2}
s!\,\varPhi_s(1,\dots,1;z_1,\dots,z_s),
\end{multline}
which can be easily found by noticing that in evaluating the residues
one has to differentiate only the factor $\prod_{j<k}(w_j-w_k)^2$.

Finally, the desired result for the emptiness formation
probability amounts in proving the identity:
\begin{multline}\label{identity2}
s! \Asym_{z_1,\dots,z_s}
\Bigg[\varPhi_s(1,\dots,1;z_1,\dots,z_s)
\prod_{1\leq j < k \leq s} [ z_j (t^2 z_j z_k - 2\Delta t z_k+1)]
\Bigg]
\\
=\frac{(-1)^{s(s+1)/2}}{\prod_{j=1}^s (z_j-1)}
\Asym_{z_1,\dots,z_s}\Bigg[
\prod_{1\leq j < k \leq s}\frac{[(t^2-2\Delta t)z_j+1](t^2 z_j z_k -2
\Delta t z_k+1)}{(z_j-1)}
\Bigg].
\end{multline}
This identity has to be used
together with identity \eqref{identity1} to reproduce \eqref{efpMIR}.
We find identity \eqref{identity2} rather difficult to
prove directly, and presently we have only been able to verify it through
computer-aided calculations for small values of $s$.
We note that rather similar identities
have been discussed in \cites{DZ-08,Z-07,TW-08}.

In conclusion, in this paper we have introduced and calculated a nonlocal
correlation function of the six-vertex model with domain wall boundary
conditions, the row configuration probability. It is given as a
product of two factors which can be treated as the partition functions on
upper and lower sublattices of the original lattice (see figure
\ref{fig.corr}). We have represented these partition functions in terms of
multiple integrals, see \eqref{oldrep_mir} and \eqref{MIRZ2}. The row
configuration probability can be used for computing other correlation
functions, provided that sums like those appearing in \eqref{efp} can be
evaluated. To illustrate this, we have considered the problem of reproducing
the known result for the emptiness formation probability. We have shown that
in this case the problem boils down to identity \eqref{identity2}. A direct
proof  of this identity, in addition to the indirect one following from the
known equality of \eqref{efpMIR} and \eqref{efpdoubleMIR}, could be useful
for the evaluation of other correlation functions.

%%%%%%%%%%%%%%%%%%%%%%%%%%%%%%%%%%%%%%%%%%%%%%%%%%%%%%%%%%%%%%%%%%%%%%%%
\section*{Acknowledgments}

We are indebted to N. M. Bogoliubov, D. Romik, and P. Zinn-Justin for useful
discussions. F.C. acknowledges partial support from MIUR, PRIN
grant 2007JHLPEZ, and from the European Science Foundation programme INSTANS.
A.G.P. acknowledges financial support from the Alexander von
Humboldt Foundation, during his stay in the University of Wuppertal.
A.G.P. also acknowledges partial support from INFN, Sezione di Firenze, from the
Russian Foundation for Basic Research (grant 10-01-00600), and from the
Russian Academy of Sciences programme ``Mathematical Methods in Nonlinear
Dynamics''.

\bibliography{mpc_bib}

\begin{bibdiv}
\begin{biblist}

\bib{KBI-93}{book}{
      author={Korepin, V.~E.},
      author={Bogoliubov, N.~M.},
      author={Izergin, A.~G.},
       title={Quantum inverse scattering method and correlation functions},
   publisher={Cambridge University Press},
     address={Cambridge},
        date={1993},
}

\bib{K-82}{article}{
      author={Korepin, V.~E.},
       title={Calculations of norms of {B}ethe wave functions},
        date={1982},
     journal={Commun. Math. Phys.},
      volume={86},
       pages={391\ndash 418},
}

\bib{I-87}{article}{
      author={Izergin, A.~G.},
       title={Partition function of the six-vertex model in the finite volume},
        date={1987},
     journal={Sov. Phys. Dokl.},
      volume={32},
       pages={878\ndash 879},
}

\bib{ICK-92}{article}{
      author={Izergin, A.~G.},
      author={Coker, D.~A.},
      author={Korepin, V.~E.},
       title={Determinant formula for the six-vertex model},
        date={1992},
     journal={J. Phys. A},
      volume={25},
       pages={4315\ndash 4334},
}

\bib{CP-08}{article}{
      author={Colomo, F.},
      author={Pronko, A.~G.},
       title={The limit shape of large alternating-sign matrices},
        date={2010},
     journal={SIAM J. Discrete Math.},
      volume={24},
       pages={1558\ndash 1571},
      eprint={0803.2697},
}

\bib{CP-09}{article}{
      author={Colomo, F.},
      author={Pronko, A.~G.},
       title={The arctic curve of the domain-wall six-vertex model},
        date={2010},
     journal={J. Stat. Phys.},
      volume={138},
       pages={662\ndash 700},
      eprint={0907.1264},
}

\bib{BKZ-02}{article}{
      author={Bogoliubov, N.~M.},
      author={Kitaev, A.~V.},
      author={Zvonarev, M.~B.},
       title={Boundary polarization in the six-vertex model},
        date={2002},
     journal={Phys. Rev. E},
      volume={65},
       pages={026126},
      eprint={cond-mat/0107146},
}

\bib{BPZ-02}{article}{
      author={Bogoliubov, N.~M.},
      author={Pronko, A.~G.},
      author={Zvonarev, M.~B.},
       title={Boundary correlation functions of the six-vertex model},
        date={2002},
     journal={J. Phys. A},
      volume={35},
       pages={5525\ndash 5541},
      eprint={math-ph/0203025},
}

\bib{FP-04}{article}{
      author={Foda, O.},
      author={Preston, I.},
       title={On the correlation functions of the domain wall six vertex
  model},
        date={2004},
     journal={J. Stat. Mech.},
      volume={0411},
       pages={P001},
      eprint={math-ph/0409067},
}

\bib{CP-05c}{article}{
      author={Colomo, F.},
      author={Pronko, A.~G.},
       title={On two-point boundary correlations in the six-vertex model with
  domain wall boundary conditions},
        date={2005},
     journal={J. Stat. Mech. Theory Exp.},
       pages={P05010},
      eprint={math-ph/0503049},
}

\bib{CP-07b}{article}{
      author={Colomo, F.},
      author={Pronko, A.~G.},
       title={Emptiness formation probability in the domain-wall six-vertex
  model},
        date={2008},
     journal={Nucl. Phys. B},
      volume={798 [FS]},
       pages={340\ndash 362},
      eprint={0712.1524},
}

\bib{M-11}{article}{
      author={Motegi, K.},
       title={Boundary correlation functions of the six and nineteen vertex
  models with domain wall boundary conditions},
      eprint={1101.0187},
}

\bib{DR-09}{article}{
      author={Francesco, P.~Di},
      author={Reshetikhin, N.},
       title={Asymptotic shapes with free boundaries},
        date={2009},
      eprint={0908.1630},
}

\bib{FR-10}{article}{
      author={Fischer, I.},
      author={Romik, D.},
       title={More refined enumerations of alternating sign matrices},
        date={2009},
     journal={Adv. Math.},
      volume={222},
      number={6},
       pages={2004\ndash 2035},
      eprint={0903.5073},
}

\bib{TF-79}{article}{
      author={Takhtadjan, L.~A.},
      author={Faddeev, L.~D.},
       title={The quantum method of the inverse problem and the {H}eisenberg
  {XYZ} model},
        date={1979},
     journal={Russ. Math. Surveys},
      volume={34},
      number={5},
       pages={11\ndash 68},
}

\bib{IKR-87}{article}{
      author={Izergin, A.~G.},
      author={Korepin, V.~E.},
      author={Reshetikhin, N.~Yu.},
       title={Correlation functions in a one-dimensional {B}ose gas},
        date={1987},
     journal={J. Phys. A},
      volume={20},
       pages={4799\ndash 4822},
}

\bib{B-82}{book}{
      author={Baxter, R.~J.},
       title={Exactly solved models in statistical mechanics},
   publisher={Academic Press},
     address={San Diego, CA},
        date={1982},
}

\bib{KMT-99}{article}{
      author={Kitanine, N.},
      author={Maillet, J.-M.},
      author={Terras, V.},
       title={Form factors of the {XXZ} {H}eisenberg spin-$1/2$ finite chain},
        date={1999},
     journal={Nucl. Phys. B},
      volume={554},
       pages={647\ndash 678},
      eprint={math-ph/9907019},
}

\bib{F-06}{article}{
      author={Fischer, I.},
       title={The number of monotone triangles with prescribed bottom row},
        date={2006},
     journal={Adv. in Appl. Math.},
      volume={37},
      number={2},
       pages={249\ndash 267},
      eprint={math/0501102},
}

\bib{S-75}{book}{
      author={Szeg{\"o}, G.},
       title={Orthogonal polinomials},
     edition={4},
      series={American Colloquium Publications},
   publisher={American Mathematical Society},
     address={Providence, RI},
        date={1975},
      volume={XXIII},
}

\bib{Zj-00}{article}{
      author={Zinn-Justin, P.},
       title={Six-vertex model with domain wall boundary conditions and
  one-matrix model},
        date={2000},
     journal={Phys. Rev. E},
      volume={62},
       pages={3411\ndash 3418},
      eprint={math-ph/0005008},
}

\bib{DZ-08}{article}{
      author={Zinn-Justin, P.},
      author={Francesco, P.~Di},
       title={Quantum {K}nizhnik-{Z}amolodchikov equation, totally symmetric
  self-complementary plane partitions, and alternating sign matrices},
        date={2008},
     journal={Theor. Math. Phys.},
      volume={154},
      number={3},
       pages={331\ndash 348},
      eprint={math-ph/0703015},
}

\bib{Z-07}{misc}{
      author={Zeilberger, D.},
       title={Proof of a conjecture of {P}hilippe {D}i {F}rancesco and {P}aul
  {Z}inn-{J}ustin related to the q{KZ} equation and to {D}ave {R}obbins' two
  favorite combinatorial objects},
        date={2007},
  url={http://www.math.rutgers.edu/~zeilberg/mamarim/mamarimhtml/diFrancesco.html},
}

\bib{TW-08}{article}{
      author={Tracy, C.~A.},
      author={Widom, H.},
       title={Integral formulas for the asymmetric simple exclusion process},
        date={2008},
     journal={Commun. Math Phys.},
      volume={279},
       pages={815\ndash 844},
      eprint={0704.2633},
}

\end{biblist}
\end{bibdiv}
\end{document}